\documentclass[epj]{svjour}
\usepackage{graphics}
\usepackage{amssymb}
\usepackage{epsfig}
\usepackage{axodraw}
\usepackage{amsmath}
\begin{document}
\title{Can the total angular momentum of $u$-quarks in the nucleon
be accessed at {\bf {\sc Hermes}}? }
\author{F.~Ellinghaus\inst{1} \and W.-D.~Nowak\inst{2} \and
A.V.~Vinnikov\inst{2,3} \and
Z.~Ye\inst{4}
}                     
%
%
\institute{Department of Physics, University of Colorado, Boulder,
Colorado 80309-0390, USA
\and DESY, D-15738 Zeuthen, Germany
\and BLTP JINR, 141980, Dubna, Moscow region, Russia
\and DESY, D-22603 Hamburg, Germany
}
\date{Received: date / Revised version: date}
%
\abstract{
We investigate the possibility to acquire information on the generalized
parton distribution $E$
and, through a model for $E$, also on the $u$-quark total angular 
momentum $J_u$ by studying deeply virtual Compton scattering
and hard exclusive $\rho^0$ electroproduction 
on a transversely polarized hydrogen target at {\sc Hermes}.
It is found that a change in $J_u$ from zero to 0.4
corresponds to a 4$\sigma$ (2$\sigma$)
difference in the calculated
transverse target-spin asymmetry
in deeply virtual Compton scattering ($\rho^0$
electroproduction), where $\sigma$ is the total experimental uncertainty.
} 
\authorrunning {F.~Ellinghaus {\it et al.}}
\title{Can the angular momentum of $u$-quarks in the nucleon be accessed
at {\sc Hermes}? }
\maketitle
%
%
%
%

\section{Introduction}
\label{intro}

\begin{figure}
\centering
\begin{picture}(150,110)(0,-5)
\ArrowLine(40,25)(50,85)
\ArrowLine(100,85)(110,25)
\SetWidth{2.0}
\ArrowLine(5,15)(40,25)
\ArrowLine(110,25)(145,15)
\Text(5,70)[]{{\large $(x+\xi)\frac{p_1+p_2}{2}$}}
\Text(145,70)[]{{\large $(x-\xi)\frac{p_1+p_2}{2}$}}
\Text(5,1)[]{{\large $p_1$}}
\Text(145,1)[]{{\large $p_2$}}
\COval(75,25)(10,35)(0){Gray}{Gray}
\end{picture}
\caption{In the parton picture, GPDs describe correlations between two
partons with different longitudinal momenta at given $Q^2$ and $t$, where
$t=(p_1-p_2)^2$ also contains transverse degrees of freedom.}
\label{kinemat1}
\end{figure}
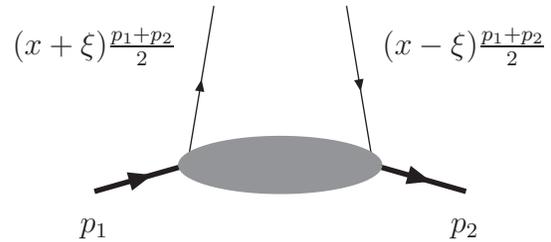

Over more than two decades, inclusive and semi-inclusive
charged lepton scattering has been used as a~powerful tool to
successfully study the longitudinal momentum structure of the nucleon,
which was parameterized in terms of parton distribution functions (PDFs).
Hard exclusive reactions can be described in
the theoretical framework of generalized parton distributions (GPDs)
\cite{gpds1,gpds2,gpds3,gpds4,gpds5}. Their
application became apparent after it had been shown \cite{jirule} that
measurements of the second moment of the sum of the `unpolarized' GPDs $H$
and $E$ open, for the first time,
access to the total angular momentum of partons in the nucleon:
\begin{equation}
J_a(Q^2)=\frac{1}{2}\underset{t\to 0}{\lim}\int\limits_{-1}^1 x\Bigl
[ H_a (x,\xi,t,Q^2) + E_a(x,\xi,t,Q^2) \Bigr ]dx.
\label{jirel}
\end{equation}
In this relation $H_a (x,\xi,t,Q^2)$ and $E_a(x,\xi,t,Q^2)$ denote parton
spin non-flip and spin flip GPDs ($a=u,d,s$),
respectively\footnote{Throughout this paper the GPD definitions of a recent
review \cite{markus} are used.}.
GPDs depend on the fractions $x$ and $\xi$ of longitudinal momentum
of the proton carried by the parton and on $t=(p_1 - p_2)^2$,
the square of the 4-momentum transfer between initial and final
protons (see. Fig.~\ref{kinemat1}). As ordinary PDFs, also
GPDs are subject to QCD evolution. Their $Q^2$ dependence has
been perturbatively calculated
up to next-to-leading order in $\alpha_s$  \cite{belevol}
and is omitted in the notations throughout
the paper.

Recently, a simultaneous description
of the transverse spatial and the longitudinal momentum structure of
the nucleon was shown to be an appealing interpretation of GPDs
\cite{impact1,impact3,impact4,impact2}.
The concept of GPDs covers several types of processes,
ranging from inclusive deeply inelastic lepton scattering to hard exclusive
Compton scattering and meson production. Measurements of GPDs are expected
to shed light especially on the hitherto theoretically uncharted territory
of long-range (`soft') phenomena where parton-parton correlations are
known to play an important role.

First steps towards the extraction of the GPD $H$ have already been
performed by scattering leptons off unpolarized protons through
measurements of either cross sections \cite{h1dvcs,zeusdvcs}, or cross
section asymmetries
with respect to beam charge \cite{hermesdvcs} or beam spin
\cite{clasdvcs1,clasdvcs2}.
Future measurements of the transverse target-spin asymmetry (TTSA) in
hard exclusive electroproduction of a real photon (deeply virtual
Compton scattering, DVCS)
or a vector meson offer the possibility to acquire
information on the spin-flip GPD $E$.
The most promising experiments to access
it are those running at intermediate energy,
where the spin-flip amplitude is expected to be sizable, while
at higher energies it is suppressed due to $s$-channel
helicity conservation. Thus at present a realistic
program may be envisaged for {\sc Hermes}, {\sc Clas}
and possibly {\sc Compass}. In this paper the prospects
are discussed for {\sc Hermes} measurements of TTSAs
in DVCS and $\rho^0$ electroproduction, and in particular 
their sensitivity to the $u$-quark total angular momentum.

\section{Modeling Generalized Parton Distributions}
\label{modelgpds}
GPDs are most commonly parameterized using an ansatz based
on double distributions \cite{radyushkin,musatov} complemented with the
D-term \cite{dterm1}. Factorizing out the $t$-dependence,
the non-forward GPDs can be related to the ordinary PDFs
and the proton elastic form factors. In this framework \cite{gpv},
the spin non-flip GPD $H$ is given by
\begin{equation}
H_{q,g}(x,\xi,t)=\frac{1-(1+\kappa_p) t/4m^2}{1-t/4m^2}\frac{H_{q,g}(x,\xi)}
{(1-t/0.71)^2},
\label{factansatz}
\end{equation}
where  $\kappa_p$=1.793 is the proton anomalous magnetic moment
and $m$ is the proton mass.
The neutron Dirac form factor is neglected compared
to the one of the proton.

For quarks, the $t$-independent part of the GPDs $H_q$ is written as
\begin{equation}
H_{q}(x, \xi)\,=\,H_q^{DD}(x, \xi) \,+\,
\theta(\xi-|x|) D_q\left ( \frac{x}{\xi} \right ) ,
\label{dtermh}
\end{equation}
where $D_q \left ( \frac{x}{\xi}\right ) $ is the D-term,
and $H_{q}^{DD}$ is the part of the GPD that is obtained
from the double distribution (DD) $F_{q}$:
\begin{equation}
H_q^{DD}(x,\xi)=
\int\limits_{-1}^{1}d\beta\
\int\limits_{-1+|\beta|}^{1-|\beta|} d\alpha\
\delta(x-\beta-\alpha\xi)\  F_q(\beta,\alpha)\ .
\label{factans}
\end{equation}
For the double distributions the suggestion of Ref.~\cite{radyushkin}
is used,
\begin{equation}
F_q(\beta, \alpha)=h(\beta,\alpha) q(\beta),
\end{equation}
where the profile function is given by \cite{musatov}:
\begin{equation}
h(\beta, \alpha) = \frac{\Gamma(2b+2)}{2^{2b+1}\Gamma^2(b+1)}\
\frac{\bigl[(1-|\beta|)^2-\alpha^2\bigr]^{b}}{(1-|\beta|)^{2b+1}} .
\label{quarkdd}
\end{equation}
For $\beta>0$, $q(\beta)=q_{val}(\beta)+\bar q(\beta) $
is the ordinary quark density for the flavor $q$.
The negative $\beta$ range corresponds to the antiquark density:
$q(-\beta)=-{\bar q}(\beta)$.
The parameter $b$ characterizes to what extent the GPD depends
on the skewness $\xi$. In the limit $b\to\infty$ the GPD
is independent on $\xi$, {\it i.e.}, $H(x,\xi)=q(x)$. Note that
$b$ is a free parameter for valence quarks ($b_{val}$) or
sea quarks ($b_{sea}$) and thus can be used as a fit parameter
in the extraction of GPDs from hard electroproduction data \cite{frank}.

For gluons, the $t$-independent part of the GPD $H_g$ 
is directly given by the double distribution,
\begin{eqnarray}
&&H_g(x,\xi)=H_g^{DD}(x,\xi)= \nonumber \\
&&\int\limits_{-1}^{1}d\beta\
\int\limits_{-1+|\beta|}^{1-|\beta|} d\alpha\
\delta(x-\beta-\alpha\xi)\  \beta F_g(\beta,\alpha)
\label{factgluons}
\end{eqnarray}
with the same form of the profile function in the double distribution
\begin{equation} 
F_g(\beta,\alpha) = h(\beta,\alpha) g(\beta).
\end{equation}
The $t$-dependence for gluons is taken to be the same
as that for quarks.

The factorized ansatz (\ref{factansatz}) is the simplest way of modeling GPDs.
However, experimental studies of elastic diffractive
processes indicate that the $t$-dependence of the
cross section is entangled with its dependence on the photon-nucleon
invariant mass \cite{collins}.
Recent evidence comes from lattice QCD calculations
\cite{latticereg1,latticereg2}
and phenomenological considerations \cite{markusreg,vandreg}.
The non-factorized ansatz can be based
on soft Regge-type parameterizations.
In this case, the $t$-dependence is not factorized out and not
controlled by a
form factor as in Eq.~(\ref{factansatz}). Instead, it is kept in
Eqs.~(\ref{dtermh}), (\ref{factans}) and (\ref{factgluons}).
The $t$-depen\-dence
of double distributions is then modeled as \cite{gpv}:
\begin{equation}
F_{q,g}(\beta,\alpha,t)=F_{q,g}(\beta,\alpha)\frac{1}{|\beta|^{\alpha' t}}~,
\label{regform}
\end{equation}
which is referred to as Regge ansatz in the following.
Here $\alpha '$ is the slope of the Regge trajectory,
$\alpha_q ' =0.8$ GeV$^{-2}$ for quarks and
$\alpha_g ' = 0.25$ GeV$^{-2}$ for gluons.

In the factorized ansatz
the spin-flip quark GPDs $E_q$ are given by \cite{gpv}:
\begin{equation}
E_q(x,\xi,t)=\frac{E_q(x,\xi)}{(1-t/0.71)^2}.
\end{equation}
In the Regge ansatz the $t$--dependence is modeled in analogy to 
Eq.~(\ref{regform}).

The $t$-independent part is parameterized using the double
distribution ansatz:
\begin{equation}
E_q(x,\xi)=E_q^{DD}(x, \xi) - {\theta(\xi-|x|)}D_q\left(\frac{x}{\xi}\right).
\label{dterme}
\end{equation}
Note that the $D$-term has the same size, but the opposite sign
in Eqs.~(\ref{dterme}) and (\ref{dtermh}). Therefore, it drops out
when calculating $J_q$ according to Eq.~(\ref{jirel}).

The double distribution has a form analogous to the
spin-nonflip case:
\begin{equation}
E_q^{DD}(x,\xi)=
\int\limits_{-1}^{1}d\beta
\int\limits_{-1+|\beta|}^{1-|\beta|} d\alpha\
\delta(x-\beta-\alpha\xi)\  K_q(\beta,\alpha)\ \,
\end{equation}
with:
\begin{equation}
K_q(\beta, \alpha) = h(\beta,\alpha) e_q(\beta).
\label{quarkddflip}
\end{equation}

The spin-flip parton densities $e_q(x)$ can not be extracted from 
deep-inelastic scattering (DIS) data, unlike the case of spin non-flip
ones. Based on the chiral
quark soliton model \cite{gpv}, the spin-flip density
is taken as a sum of valence and sea quarks contributions. Since
in this model the sea part was found to be
very narrowly peaked around $x=0$, the whole density is written as:
\begin{equation}
e_q(x) = A_q q_{val}(x) + B_q \delta(x) .
\label{etotparam}
\end{equation}
In this expression, the shape of the valence quark part is
given by that of the spin non-flip density.
The coefficients $A_q$ and $B_q$ are constrained by the total angular
momentum sum rule (\ref{jirel})
and the normalization condition
\begin{equation}
\int\limits_{-1}^{+1}dx\; e_q(x)\, =\, \kappa_{q}\; ,
\label{anommoment}
\end{equation}
where $\kappa_q$ is the anomalous magnetic moment of quarks
of flavor $q$
($\kappa_u=2\kappa_p+\kappa_n=1.67$, $\kappa_d=\kappa_p+2\kappa_n=-2.03$).
The constraints yield:
\begin{eqnarray}
A_q &=& \frac{ 2 \, J_q \,-\, M^{(2)}_q} {M^{(2)}_{q_{val}}}, \\
B_u &=& 2 \left[ \frac{1}{2} \kappa_u - 
\frac{2 J_u - M^{(2)}_u}{M^{(2)}_{u_{val}}} \right]\, , \\
B_d &=& \kappa_d - 
\frac{2 J_d - M^{(2)}_d}{M^{(2)}_{d_{val}}} .
\end{eqnarray}
Here $M_q^{(2)}$ and $M^{(2)}_{q_{val}}$
are the parton momentum contributions
to the proton momentum:
\begin{equation}
M^{(2)}_{q_{val}}=\int\limits_0^1 x q_{val}(x) dx,~~~
M^{(2)}_q=\int\limits_0^1 x \left [ q_{val}(x)+2{\bar q}(x)\right ] dx.
\end{equation}
In the given scenario the total angular momenta carried by
$u$- and $d$-quarks,
$J_u$ and $J_d$, enter directly as free parameters in the parameterization
of the spin-flip GPD $E_q (x,\xi,t)$.
Hence the parameterization (\ref{etotparam}) 
can be used to investigate the sensitivity of hard electroproduction 
observables to variations in $J_u$ and $J_d$. 

As to the gluons, there exists no hint how the spin-flip GPD $E_g$ could 
be described. There is an expectation that $E_g$ is not large compared
to $E_u$ and $E_d$ \cite{ournew}.
Hence, for simplicity throughout the present
study $E_g$ is neglected (``passive'' gluons, {\it i.e.} $E_g=0$).

\begin{figure}
\centering
\epsfig{file=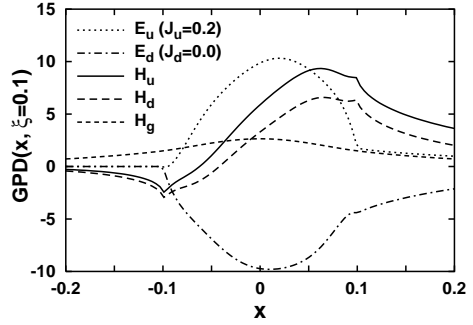,width=0.75\hsize}
\caption{$t$-independent part of quark and gluon GPDs at
$Q^2$=4~GeV$^2$, $\xi$=0.1 (MRST98 PDFs are used).}
\label{gpd1}
\end{figure}

As an example,  Fig.~\ref{gpd1} shows the $t$-independent part of various
GPDs at $\xi=0.1$, based on the MRST98 \cite{mrst98} parameterization of 
PDFs at $Q^2$=4 GeV$^2$. Using instead CTEQ6L PDFs \cite{cteq}
as input, the results for $u(d)$ quark GPDs are changed
by less than 3\%(10\%); the GPD $H_g$ is up to 40\% larger at $x=0$.
Because of $u$-quark dominance in electroproduction, uncertainties
originating from $d$-quark PDFs can be safely neglected. Since gluons
are absent in leading-order DVCS,
uncertainties resulting from gluon
PDFs are of little influence for DVCS asymmetries and have been found to 
lead to a 
fractional change  of up to 15\% for the $\rho^0$ asymmetries.
For the following calculations the MRST98 PDF set is taken.

\section{Sensitivity of DVCS to the $u$-quark Total Angular Momentum}
\label{ttsadvcs}
\subsection{Cross Section and Asymmetries}
The 5-fold cross section for the process $e(k)+p(p_1){\rightarrow}e(k')+p(p_2)+\gamma(q_2)$
is given by:
\begin{equation}
\frac{d\sigma}{dx_BdQ^2dtd\phi{d}\phi_S}=\frac{\alpha_{em}^3x_By}
{16\pi^2Q^2\sqrt{1+4x_B^2m^2/Q^2}}\cdot\left|
\frac{\mathcal{T}}{e^3}\right|^2 ,
\end{equation}
where $Q^2=-q_1^2$ is the negative squared 4-momentum of the virtual photon,
$x_B={Q^2}/{(2p_1\cdot q_1)}$ is the Bjorken variable, $t=(p_1-p_2)^2$, $y={(p_1\cdot q_1)}/{(p_1\cdot k)}$, $\mathcal{T}$ denotes the photon production amplitude and $e$ is the electron charge. Since the DVCS and Bethe-Heitler (BH) processes have an identical final state, in which the photon is radiated either from a parton or from a lepton, respectively, $\mathcal{T}$ is given by the coherent sum of the BH amplitude $\mathcal{T}_{BH}$ and the DVCS amplitude $\mathcal{T}_{DVCS}$:
\begin{equation}
\left|\mathcal{T}\right|^2=
\left|\mathcal{T}_{BH}+\mathcal{T}_{DVCS}\right|^2=
\left|\mathcal{T}_{BH}\right|^2+\left|\mathcal{T}_{DVCS}\right|^2+\mathcal{I}, 
\end{equation}
in which
\begin{equation}
\mathcal{I}=\mathcal{T}_{BH}^*\mathcal{T}_{DVCS}+\mathcal{T}_{BH}\mathcal{T}_{DVCS}^*
\end{equation}
describes the interference between both processes. 

\begin{figure}[t]
\centering
\epsfig{file=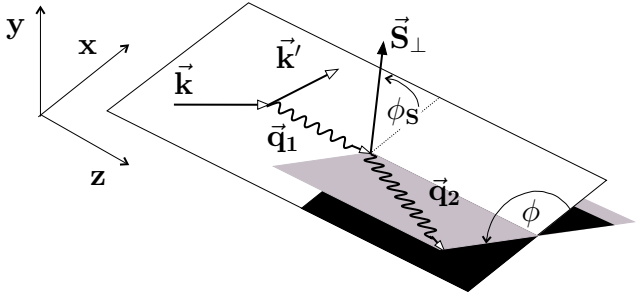,angle=270,width=0.95\hsize}
\caption{Kinematics and azimuthal angles of photon electroproduction in 
the target rest frame. The $z$-direction is chosen
along the three-momentum of the virtual photon $\vec{q}_1$. The lepton 
three-momenta $\vec{k}$ and $\vec{k}'$ form the lepton scattering plane, 
while the three-momenta of virtual and real photons $\vec{q}_1$ and $\vec{q}_2$
define the production plane. The azimuthal angle of the production plane
with respect to the scattering plane, around the virtual
photon direction, is denoted as $\phi$. Correspondingly, $\phi_S$ denotes the 
azimuthal angle of the target polarization vector with respect to the lepton 
scattering plane. In this frame the target polarization vector is given
as $\vec{S}_\perp = (\cos{\phi_S}, \sin{\phi_S}, 0)$. The definitions conform 
with the Trento conventions~\cite{Trento}.}
\label{fig:coordinate}
\end{figure}

The coordinate system is defined in the target rest frame, as
explained in Fig.~\ref{fig:coordinate}. 
The theoretical formulae used below refer to the target being transversely polarized w.r.t. the virtual photon direction, while in the experiment the target polarization is transverse w.r.t. the incident lepton direction. At {\sc Hermes} kinematics, these two directions are approximately parallel and the small longitudinal component ($<10\%$) of the target polarization 
along the virtual photon direction can be neglected. Thus the reasonable approximation
\begin{equation}
d\sigma=d\sigma_{unp}+d\sigma_{TP}
\end{equation}
is used, where $d\sigma_{unp}$ ($d\sigma_{TP}$) denotes the cross section
for the unpolarized (transversely polarized) component.

Since in the kinematic region of the {\sc Hermes} experiment the DVCS 
cross section is typically much smaller than the BH cross section 
\cite{kornow}, the contribution of the DVCS term to the total cross section
is neglected in the following. The
contributions of the BH term for 
an unpolarized beam are:
\begin{eqnarray}
\left|\mathcal{T}^{BH}_{unp}\right|^2&=& \frac{e^6}
{x_B^2y^2(1+4x_B^2m^2/Q^2)^2tP_1(\phi)P_2(\phi)} \times \nonumber\\
& & \left[c_{0,unp}^{BH}+c_{1,unp}^{BH}\cos{\phi}+c_{2,unp}^{BH}
\cos{2\phi}\right] ,\label{eqn:cs1} \\
\left|\mathcal{T}^{BH}_{TP}\right|^2&=&0 .\nonumber
\end{eqnarray}
The full expressions for the BH propagators $P_1(\phi)$, $P_2(\phi)$ and for
the Fourier coefficients $c_{i,unp}^{BH}$ can be 
found in Ref.~\cite{belitsky}\footnote{The azimuthal angles defined in
this work are different from those used in Ref.~\cite{belitsky}:
$\phi=\pi-\phi_{{\mbox{\cite{belitsky}}}}$ and $\phi-\phi_S=\pi+
\varphi_{{\mbox{\cite{belitsky}}}}$.}.

The leading twist and leading order $\alpha_s$ contributions of the DVCS-BH
interference term to the total 
cross section can be written as:
\begin{eqnarray}
&&\mathcal{I}_{unp}=\frac{\pm e^6}{x_By^3tP_1(\phi)P_2(\phi)}\left(c_{0,unp}^{I}+c_{1,unp}^{I}\cos{\phi} \right) ,\nonumber \\
&&\mathcal{I}_{TP}=\frac{\pm e^6}{x^2_By^2tP_1(\phi)P_2(\phi)}
\ f\left(x_B,y,Q^2\right) \times \label{eqn:cs2} \\
&&\Bigl [ Im\widehat{M}_N\sin{(\phi-\phi_S)}\cos{\phi}+Im\widehat{M}_S\cos{(\phi-\phi_S)}\sin{\phi} \Bigr ] .\nonumber
\end{eqnarray}
Here $+$ ($-$) stands for a negatively (positively) charged lepton beam and $f\left(x_B,y,Q^2\right)$ is a kinematic pre-factor independent of azimuthal angles. The full expressions for $c_{i,unp}^{I}$ can be found in Eqs.~(53-56) of Ref. \cite{belitsky}. $\widehat{M}_N$ and $\widehat{M}_S$ are certain linear combinations of the Compton form factors ${\mathcal{H}}$,
${\mathcal{E}}$, ${\mathcal{\widetilde{H}}}$ and ${\mathcal{\widetilde{E}}}$, 
which are convolutions of the respective twist-2 GPDs $H$, $E$, $\widetilde{H}$ and $\widetilde{E}$ with the hard-scattering kernels as defined in Eq.~(9) of Ref. \cite{belitsky}.

The full expressions for $\widehat{M}_N$ and $\widehat{M}_S$ can be found in
Eq.~(71) in Ref.~\cite{belitsky}
or in Eq.~(60) in Ref.~\cite{markusapet}. 
Since $\xi\simeq x_B/(2-x_B)$ is small in a wide 
range of experimentally relevant kinematics, terms with pre-factor $\xi$ 
or $x_B$ can be neglected, except for the GPD $\widetilde{E}$ because the pion pole contribution to $\widetilde{E}$ scales like $\xi^{-1}$, so that $\widehat{M}_N$ and $\widehat{M}_S$ can be approximated as:
\begin{eqnarray}
\widehat{M}_N&\simeq&-\frac{t}{4M^2}\cdot
\left[F_2\mathcal{H}-F_1\mathcal{E}\right],\nonumber\\
\widehat{M}_S&\simeq&-\frac{t}{4M^2}\cdot
\left[F_2\mathcal{\widetilde{H}}-F_1\xi\mathcal{\widetilde{E}}\right].\label{eqn:ref9}
\end{eqnarray}
Here $F_1$ and $F_2$ are the Dirac and Pauli form factors of the proton, respectively. 

In order to constrain the GPDs involved in Eq.~(\ref{eqn:ref9}), the transverse polarization component of the interference term, $\mathcal{I}_{TP}$, has to be singled out. This can be accomplished by forming the transverse (T) target-spin asymmetry with unpolarized (U) beam:
\begin{eqnarray}
A_{UT}(\phi&-&\phi_S)=\frac
{d\sigma(\phi-\phi_S) - d\sigma(\phi-\phi_S+\pi)}
{d\sigma(\phi-\phi_S) + d\sigma(\phi-\phi_S+\pi)} \label{eqn:tta}\\
&\simeq & A_{UT}^{\sin(\phi-\phi_S)\cos{\phi}}\cdot \sin(\phi-\phi_S)
\cos{\phi} \nonumber \\
& & +A_{UT}^{\cos(\phi-\phi_S)\sin{\phi}}
\cdot \cos(\phi-\phi_S)\sin{\phi} \label{eqn:tta10}.
\end{eqnarray}
As $\left|\mathcal{T}^{BH}_{unp}\right|^2$ and
$\mathcal{I}_{unp}$ are independent on $\phi-\phi_S$, they
do not appear in the numerators of Eq.~(\ref{eqn:tta}). Since their dominant contribution to the denominator in Eq.~(\ref{eqn:tta}) is given by $c_{0,unp}^{BH}$, the two amplitudes of the TTSA, $A_{UT}^{\sin(\phi-\phi_S)\cos{\phi}}$ and $A_{UT}^{\cos(\phi-\phi_S)\sin{\phi}}$, can be approximated as: 
\begin{eqnarray}
\label {asyms}
A_{UT}^{\sin{(\phi-\phi_S)}\cos{\phi}}&\simeq&\pm f\left(x_B,y,Q^2\right)
\cdot\frac{Im \widehat{M}_N}{c_{0,unp}^{BH}}, \nonumber \\
A_{UT}^{\cos{(\phi-\phi_S)}\sin{\phi}}&\simeq&\pm f\left(x_B,y,Q^2\right)
\cdot\frac{Im \widehat{M}_S}{c_{0,unp}^{BH}}.
\label{eqn:ref1}
\end{eqnarray}

Note that the approximations used in this section are for illustrative
purposes only and
are not used in the numerical calculations described below.

\subsection{Expected Value of TTSA and Projected Statistical Uncertainty}
\label{ttsadvcsexp}
Since in the DVCS process the gluons enter only in NLO in $\alpha_S$, 
their contributions to cross section and TTSA are neglected. 
For the quarks, it can be seen from Eq.~(\ref{eqn:ref9}) that, 
besides the GPDs $H$ and $E$ which have been discussed in 
Sect.~\ref{modelgpds}, 
there are two other GPDs, $\widetilde{H}$ and $\widetilde{E}$, 
involved in the TTSA for DVCS. Since they are not the main interest
of this paper, in the calculations below they are always included and 
kept unchanged. In their model description, the forward limit of the 
GPD $\widetilde{H}$ is fixed by the 
quark helicity distributions $\Delta q(x,\mu^2)$, 
while the GPD $\widetilde{E}$ is evaluated from the 
pion pole which only provides a real part to
$\widehat{M}_S$ in Eq.~(\ref{eqn:ref9}). 

At present, there exists a code \cite{vgg} designed to calculate
observables in the exclusive reaction $ep \to ep\gamma$. 
It has been used (see App.~\ref{ttsadvcscalc}) to evaluate the TTSA 
arising from the DVCS-BH interference. The TTSA is calculated at 
the average kinematic values per bin in $x_B$, $Q^2$ and $t$ 
taken from a measurement 
of the beam-spin asymmetry in DVCS at
{\sc Hermes} \cite{frank} (see Tab.~\ref{tab:kinematics}).

The statistical error of an asymmetry is independent
on its size if the asymmetry itself is
small. For a single beam (target) spin
asymmetry it is obtained as:
\begin{equation}
\sigma^2_{stat}\propto\frac{1}{N}\cdot\frac{1}{P_{beam(target)}^2},
\label{eqn:stat}
\end{equation}
where $N$ is the total number of events that is
proportional to the integrated luminosity,
and $P_{beam(target)}$ is the beam (target)
polarization. The following projection is based on a 
future {\sc Hermes} data set 
of 8 million DIS events to be taken with an unpolarized positron beam and a 
transversely polarized hydrogen target. 
Using the known statistical errors of the beam-spin asymmetry measurement 
at {\sc Hermes} on an unpolarized hydrogen target (7 million DIS events, 
$P_{beam}\simeq 50\%$)
\cite{frank}, the projected statistical error for the TTSA is obtained.

\begin{table}[here]
\caption{Average kinematic values for $Q^2$, $x_B$, $-t$ bins
and statistical errors, 
taken from a measurement of the beam-spin asymmetry at {\sc Hermes}
\cite{frank}.}
\label{tab:kinematics}
\begin{center}
\resizebox{\hsize}{!}{
\begin{tabular}{|c|c|c|c|c|c|}
\hline
$Q^2$ bin (GeV$^2$) & 1.00-1.50 & 1.50-2.30 &
2.30-3.50 & 3.50-6.00 & 6.00-10.0 \\\hline
$\langle{Q^2}\rangle$ (GeV$^2$) & 1.2 & 1.8 & 2.8 & 4.4 & 7.1 \\\hline
$\langle{x_B}\rangle$ & 0.06 & 0.08 & 0.10 & 0.15 & 0.24 \\\hline
$\langle{-t}\rangle$ (GeV$^2$) & 0.07 & 0.09 & 0.12 & 0.17 & 0.24 \\\hline
$stat.~\delta{A}_{LU}^{\sin{\phi}}$ & 0.053 &
0.050 & 0.061 & 0.070 & 0.163 \\\hline
\hline
$x_B$ bin & 0.03-0.07 &
0.07-0.10 & 0.10-0.15 & 0.15-0.20 & 0.20-0.35 \\\hline
$\langle{Q^2}\rangle$ (GeV$^2$) & 1.4 & 2.2 & 3.1 & 4.5 & 6.1 \\\hline
$\langle{x_B}\rangle$ & 0.05 & 0.08 & 0.12 & 0.17 & 0.24 \\\hline
$\langle{-t}\rangle$ (GeV$^2$) & 0.08 & 0.10 & 0.12 & 0.17 & 0.22 \\\hline
$stat.~\delta{A}_{LU}^{\sin{\phi}}$ & 0.048 &
0.053 & 0.060 & 0.099 & 0.145 \\\hline
\hline
$-t$ bin (GeV$^2$) & 0.00-0.06 & 0.06-0.14 & 0.14-0.30 &
0.30-0.50 & 0.50-0.70 \\\hline
$\langle{Q^2}\rangle$ (GeV$^2$) & 2.0 & 2.5 & 3.0 & 3.6 & 3.9 \\\hline
$\langle{x_B}\rangle$ & 0.08 & 0.10 & 0.11 & 0.12 & 0.12 \\\hline
$\langle{-t}\rangle$ (GeV$^2$) & 0.03 & 0.09 & 0.20 & 0.37 & 0.57 \\\hline
$stat.~\delta{A}_{LU}^{\sin{\phi}}$ &
0.041 & 0.052 & 0.066 & 0.126 & 0.263 \\\hline
\end{tabular} }
\end{center}
\end{table}

\begin{figure}[htb]
\epsfig{file=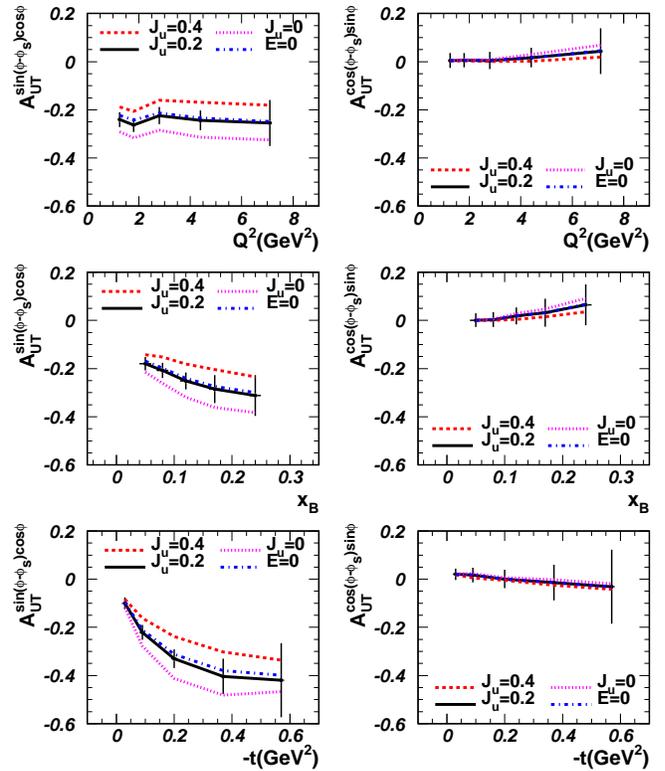,width=\hsize}
\caption{Expected DVCS TTSA amplitudes 
$A_{UT}^{\sin{(\phi-\phi_S)}\cos{\phi}}$ and
$A_{UT}^{\cos{(\phi-\phi_S)}\sin{\phi}}$ in the Regge ansatz for
$b_{val}=1$, $b_{sea}=\infty$,  $J_u=0.4$ ($0.2,0.0$), $J_d=0.0$. $E=0$ denotes
zero effective contribution from the quark GPDs $E_q$.
The calculations are done at the average kinematic values as listed
in Tab.~\ref{tab:kinematics}. Projected statistical errors are shown.
\label{fig:tta8}}
\end{figure}

\begin{figure}[htb]
\epsfig{file=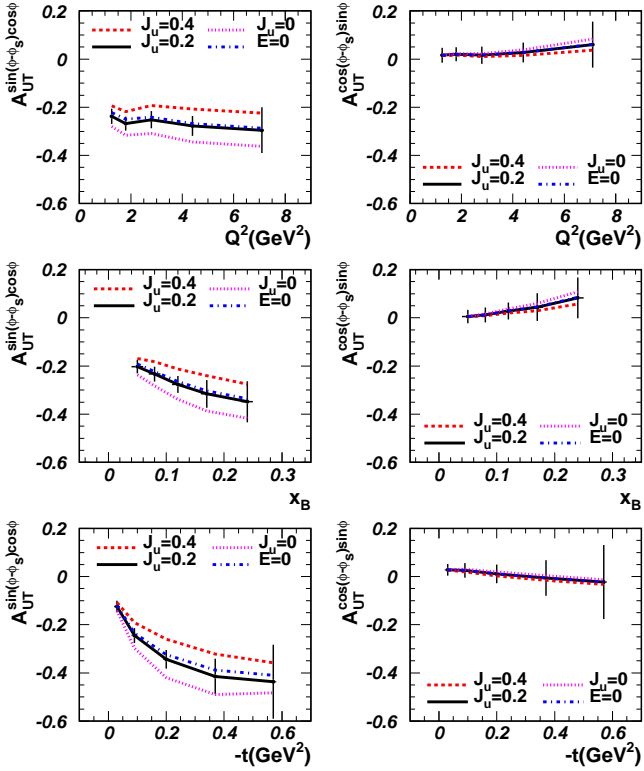,width=\hsize}
\caption{Expected DVCS TTSA amplitudes
$A_{UT}^{\sin{(\phi-\phi_S)}\cos{\phi}}$ and
$A_{UT}^{\cos{(\phi-\phi_S)}\sin{\phi}}$ in the Regge ansatz for
$b_{val}=1$, $b_{sea}=1$,  $J_u=0.4$ ($0.2,0.0$), $J_d=0.0$. $E=0$ denotes
zero effective contribution from the quark GPDs $E_q$.
The calculations are done at the average kinematic values as listed
in Tab.~\ref{tab:kinematics}. Projected statistical errors are shown.
\label{fig:tta9}}
\end{figure}

The projections for 
$A_{UT}^{\sin{(\phi-\phi_S)}\cos{\phi}}$ 
and $A_{UT}^{\cos{(\phi-\phi_S)}\sin{\phi}}$ 
are calculated
for different values of the total
angular momentum $J_u$. 
Since the contributions of $u$-quark and $d$-quark are proportional
to the corresponding squared charge, 
the $d$-quark contribution is suppressed and hence 
in the calculations a fixed value is used for $J_d$.
The latter was chosen to be $J_d=0$, 
inspired by the results of recent lattice calculations (see
e.g. Ref.~\cite {lattice_prl_qcdsf}).
Using both Regge and factorized ans\"{a}tze,
the asymmetries are calculated
for the four possible 
cases setting the profile parameters $b_{val}$ and $b_{sea}$ to either
one or infinity.
Comparing all sets of projections to each other, the amplitudes of the TTSA
appear to be sensitive only to the change in $b_{sea}$ from one to infinity. 
The resulting differences are small and can be seen by comparing 
Figs.~\ref{fig:tta8} and \ref{fig:tta9}, where the amplitudes
are shown in dependence on $Q^2$, $x_B$ and $-t$ together with the projected 
statistical errors. In order to study the contributions of the GPDs $H$, 
$\widetilde{H}$ and $\widetilde{E}$ alone, calculations are done 
for $E=0$ as well. 

As expected from Eqs.~(\ref{eqn:ref9}) and (\ref{asyms}), variations in
the parameter settings for the GPD $E$ become
manifest in $A_{UT}^{\sin{(\phi-\phi_S)}\cos{\phi}}$
while $A_{UT}^{\cos{(\phi-\phi_S)}\sin{\phi}}$ shows only minor 
modifications. The latter are apparent only in the kinematic regime
of large $x_B$ or correspondingly large $Q^2$ since the contribution
of the GPDs $E_q$ to $\widehat{M}_S$
is suppressed by $x_B$ and thus has been neglected in Eq.~(\ref{eqn:ref9}).
Within these model calculations
$A_{UT}^{\sin{(\phi-\phi_S)}\cos{\phi}}$
turns out to be sizable even when the calculation is done for $E_q=0$.
Thus a solid knowledge about the GPD $H_u$ is needed
in order to constrain $J_u$.
It has been shown \cite{dis04} that the model parameters for the GPD $H_u$,
in particular the size of the profile parameters 
$b_{val}$ and $b_{sea}$, can be well constrained by the
envisaged {\sc Hermes} DVCS measurements until 2007, using an unpolarized 
hydrogen
target. Since in addition the profile parameters are assumed to be the 
same for the GPD $E_u$, the only remaining free parameter is $J_u$. Hence 
the projected measurement of $A_{UT}^{\sin{(\phi-\phi_S)}\cos{\phi}}$ 
has a clear potential to constrain $J_u$, as can be seen from the left 
panels of Figs.~\ref{fig:tta8} and \ref{fig:tta9}.

\begin{figure}[htb]
\centering
\epsfig{file=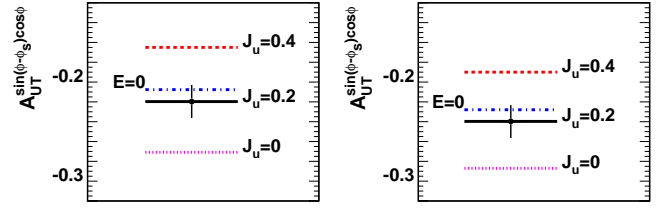,width=\hsize}
\caption{Expected DVCS TTSA amplitudes $A_{UT}^{\sin{(\phi-\phi_S)}\cos{\phi}}$
with $b_{val}=1$ and $b_{sea}=\infty$ (left panel) or $b_{sea}=1$ 
(right panel),
$J_u=0.4$ ($0.2,0.0$), $J_d=0.0$
in the Regge ansatz at the average kinematics of the full measurement.
$E=0$ denotes zero effective contribution from the GPDs $E_q$.
The projected statistical error for 8 million DIS events
is shown. The systematic error is expected to not exceed the statistical one.
\label{fig:combine}}
\end{figure}

The discriminative power of the envisaged TTSA measurement
can be enhanced by combining all data into one point, since all considered
models show the same kinematic dependences.
The corresponding statistical power of a {\sc Hermes} data set based on 8 
million 
DIS events is shown in Fig.~\ref{fig:combine}, for $b_{sea}$ equal to one 
or infinity, and for three different
values of the total $u$-quark angular momentum $J_u$ plus the special case
$E_q=0$.

It appears that for $b_{sea}=1$ ($b_{sea}=\infty$) the amplitude ranges 
between values of -0.17 and -0.27 (-0.19 and -0.29) when $J_u$ ranges 
between 0.4 and 0. The projected statistical error for 
these integrated TTSA amplitudes is 0.017. Extrapolating the knowledge
on the systematic uncertainty from the analysis
of 2000 {\sc Hermes} data \cite{frank}, its size can be expected to not exceed
the statistical error, such that a total experimental uncertainty
below 0.025 appears as a realistic estimate. 
Altogether, the difference in the size of the TTSA due to a change of $J_u$
between zero and 0.4 corresponds to a 4$\sigma$ effect, where $\sigma$
denotes the {\em total} experimental uncertainty. 
Thus, based on the GPD model used it can be expected that the upcoming
DVCS results from {\sc Hermes}\footnote{The recent switch of
the accelerator (HERA) to an electron beam
will require to also perform the above calculations for the negative beam
charge. However, the sensitivity to $J_u$ of the combined electron and
positron measurements is expected to be similar to the one calculated here
for a positron beam only.}
will provide a constraint on the size of $J_u$.

\section{Sensitivity of Elastic $\rho^0$ Electroproduction to the
$u$-quark Total Angular Momentum}
\label{ttsarho}
Also a measurement of the TTSA in elastic
vector meson electroproduction can be a source of information about the 
spin-flip generalized parton distribution $E$. An estimate for the
asymmetry was obtained in Ref. \cite{gpv}, using the factorized
model of GPDs described in Sect. 2 without inclusion of gluons.
The scope of this section is to also include the Regge ansatz,
to check the assumption that
the gluon contribution to the $\rho^0$ electroproduction cross section
is small, and to eventually calculate the size of the TTSA at {\sc Hermes} kinematics.
The issue is raised since in contrast to DVCS, in vector meson
elastic electroproduction gluons enter at the same order
of $\alpha_s$ as quarks, namely at order $\alpha_s$ to the power one.
Hence this channel appears as one of the rare cases where {\it gluon} 
GPDs may be accessed through {\sc Hermes} data.

\subsection{Cross Section and Gluonic Contribution}
\label{ttsarhocross}
It was shown \cite{frankstrik97} that the leading twist contribution
to exclusive electroproduction of vector
mesons requires both the virtual photon and the vector meson to be
longitudinal, {\it i.e.} transversely polarized.
Therefore the present calculations cover only the 
longitudinal part of the cross section.

The cross section of the reaction $\gamma^*_L p \to \rho^0 p'$
is given by \cite{gpv}
\begin{equation}
\frac{d \sigma_L}{dt}=\frac{1}{8m\pi(W^2-m^2)|{\vec q}_{1}| }
\Bigl ( |{\mathcal T}_A|^2 + |{\mathcal T}_B|^2 \Bigr ),
\end{equation}
where ${\vec q}_1$ is the momentum of the virtual
photon in the center of mass
system of this photon and the initial proton, while
$W$ is their invariant mass. The spin-flip amplitude reads
\begin{eqnarray}
{\mathcal T}_A=-i e \frac{2\sqrt{2}\pi\alpha_s}
{9 Q}{\mathcal A} {\bar u}(p_2)n^{\mu}\gamma_{\mu}u(p_1)
\int\limits_0^1 dz \frac{\Phi(z)}{z} \\
\simeq
-i {\mathcal A}
\pi e\alpha_s \frac{8}{9}\frac{1}{Q}\int\limits_0^1 dz
\frac{\Phi (z)}{z}, \nonumber
\end{eqnarray}
and the spin-flip one
is\footnote{in the subsequent calculations the exact formulae
were used}:
\begin{eqnarray}
{\mathcal T}_B=
e \frac{\sqrt{2}\pi\alpha_s}{9 Q m}
{\mathcal B} {\bar u}(p_2)\sigma^{\mu\nu}n_{\mu}\Delta_{\nu}
u(p_1) \int\limits_0^1 dz \frac{\Phi(z)}{z} \\
\simeq -i {\mathcal B}
\pi e\alpha_s \frac{|\Delta_T|}{m}\frac{4}{9}
\frac{1}{Q}\int\limits_0^1 dz \frac{\Phi (z)}{z}. \nonumber
\end{eqnarray}
Here $n=(1,0,0,-1)/(\sqrt{2}(p_1+p_2)^+)$
is a light-like vector along the $z$-axis, $\Delta=p_2-p_1$ is the
4-momentum transfer ($\Delta^2=t$). The modulus of its transverse component
is given by $|\Delta_T|=\sqrt{-t(1-\xi^2)-4\xi^2m^2}$.
The $\rho^0$-meson wave
function is taken in the form
\begin{equation}
\Phi(z)=6z(1-z)f_{\rho}
\end{equation}
with $f_{\rho}$=0.216 GeV and $z$ being the meson longitudinal momentum
fraction carried by a parton. The complex factors
${\mathcal A}$ and ${\mathcal B}$ are given by:
\begin{eqnarray}
{\mathcal A}=\frac{1}{\sqrt{2}}\int\limits_{-1}^{1}\Bigl ( e_u H_u(x,\xi,t)
&-&e_d H_d(x,\xi,t) - \nonumber \\
&&\frac{3}{8}(e_u-e_d)\frac{H_g(x,\xi,t)}{x} \Bigr ) \times
\nonumber \\
\Bigl \{ \frac{1}{x-\xi+i \epsilon}&+&\frac{1}{x+\xi-i\epsilon}
\Bigr \} dx,
\label{singintegr1}
\end{eqnarray}
\begin{eqnarray}
{\mathcal B}=\frac{1}{\sqrt{2}}\int\limits_{-1}^{1}\Bigl ( e_u E_u(x,\xi,t)
&-&e_d E_d(x,\xi,t) -\nonumber \\
&&\frac{3}{8}(e_u-e_d)\frac{E_g(x,\xi,t)}{x} \Bigr ) \times \nonumber\\
\nonumber \\
\Bigl \{ \frac{1}{x-\xi+i \epsilon}&+&\frac{1}{x+\xi-i \epsilon}
\Bigr \} dx.
\label{singintegr2}
\end{eqnarray}

The TTSA is defined as
\begin{eqnarray}
A_{UT}(\phi-\phi_S)&=&\frac
{d\sigma(\phi-\phi_S) - d\sigma(\phi-\phi_S+\pi)}
{d\sigma(\phi-\phi_S) + d\sigma(\phi-\phi_S+\pi)} \nonumber \\
&=& A_{UT}^{\sin(\phi-\phi_S)}\cdot \sin(\phi-\phi_S).
\end{eqnarray}
The $A_{UT}^{\sin(\phi-\phi_S)}$ amplitude of the TTSA
can be expressed in terms of the spin flip and spin non-flip
amplitudes as \cite{gpv}:
\begin{eqnarray} \label{ttrasyform}
&& A_{UT}^{\sin(\phi-\phi_S)} = \\
&& \frac{Im ({\mathcal A}{\mathcal B}^*)|\Delta_T|/m}
{(1-\xi^2)|{\mathcal A}|^2-(\xi^2+\frac{t}{4m^2})|{\mathcal B}|^2
-2\xi^2 Re({\mathcal A}{\mathcal B}^*)}. \nonumber
\end{eqnarray}
Note that using the Trento convention~\cite{Trento} the sign of this equation 
is opposite to that in Ref.~\cite{gpv} and the normalization is larger by
a factor of $\pi/2$.

\begin{figure}
\centering
\epsfig{file=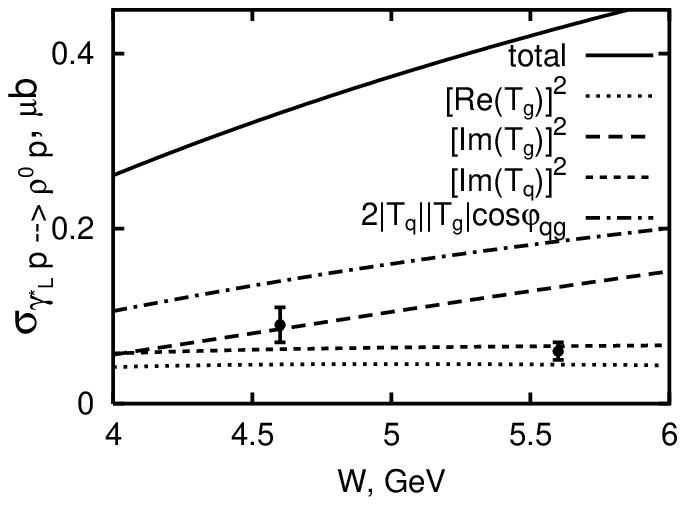,width=0.7\hsize}
\epsfig{file=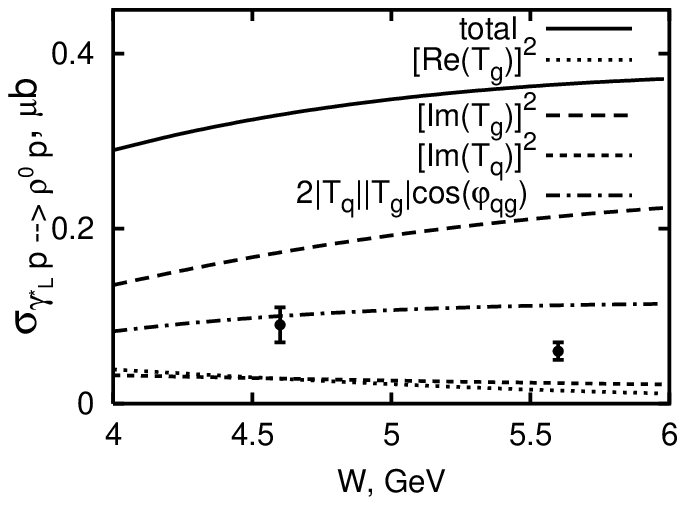,width=0.7\hsize}
\caption{The calculated $W$-dependence of the hard exclusive $\rho^0$
electroproduction cross section at $Q^2=4 $ GeV$^2$ for the factorized (top)
and Regge (bottom) GPD models compared to {\sc Hermes} data \cite{rhoherm}.
${\cal T}_q$ is the quark amplitude and ${\cal T}_g$ the gluon amplitude.
The quark real part is very small and is not shown.}
\label{rhocross}
\end{figure}

The cross section is calculated using both factorized
and Regge ans\"{a}tze for GPDs\footnote{The principal values of the
integrals in Eqs.~(\ref{singintegr1}) and (\ref{singintegr2}) are calculated
in the following way: $\int\limits_{a<0}^{b>0}
\frac{f(x)}{x} dx = f(b)\ln(b)-f(a)\ln(a) -\int\limits_a^0 f'(x)\ln(x)dx
+\int\limits_0^b f'(x)\ln(x) dx$. In this way the
non-integrable singularity is exchanged by an integrable one.}.
The value $b=1$ is taken for the profile parameter both for
sea and valence quarks.
It is found that using $b_{sea}=\infty$ instead of $b_{sea}=1$
leads to a rise of the cross section by a factor
of about $1.15$. The value $b_{val}=b_{sea}=1$
is chosen to provide a direct comparison
to previous calculations \cite{gpd,gpv}.
The value of the profile parameter for gluons is chosen
as $b$ = 2 and it has been checked that choosing $b$ = 1 or $\infty$
does not change the cross section by more than 20\%.
The $W$-dependence of the cross section for
$Q^2$=4 GeV$^2$ is shown in Fig.~\ref{rhocross}.
For both ans\"{a}tze the calculations overshoot considerably
the experimental data from {\sc Hermes} \cite{rhoherm}.
However, a significant reduction of the calculated cross section
might be expected if transverse motion effects are taken into account
\cite{gpv,gpd}.
On the other hand, also
the double distribution based calculations of the DVCS cross section
have been found to overshoot the data from H1
\cite{freundmac1,freundmac2}.

An unexpected result of the calculation shown in Fig.~\ref{rhocross}
is a quite small (15-20\%) pure quark contribution
to the cross section, while in Refs.~\cite{gpv,gpd}
the quark contribution was found to be dominant.
Comparing the calculated quark contribution to experimental data
(Fig.~\ref{rhocross}) it could also be concluded that the 
gluon contribution in the present calculation is substantially 
overestimated, while the quark contribution itself is reasonable and 
can explain alone (in the factorized ansatz) the value of
the measured cross section. However, 
there exists experimental evidence that the gluon 
spin non-flip part is indeed large
\cite{ournew}.

On the amplitude level, the cross sections of
$\rho^0$ and $\phi$ mesons are given as:
\begin{eqnarray}
\sigma_{\rho^0}&=&C_{\rho^0}|{\cal T}_q+{\cal T}_g|^2 \nonumber \\
&=&C_{\rho^0}(|{\cal T}_q|^2+ 2|{\cal T}_q||{\cal T}_g|\cos(\varphi_{qg})+
|{\cal T}_g|^2), \\
\sigma_{\phi}&=&\frac{2}{9}C_{\phi}|{\cal T}_g|^2. \nonumber
\end{eqnarray}
Here ${\cal T}_q$ is the quark amplitude and ${\cal T}_g$
the gluon amplitude (the
$s$-quark contribution to the $\phi$ production amplitude
is neglected) and $\varphi_{qg}$ is the effective phase between the quark
and gluon amplitudes.
In the existing GPD-based calculations \cite{gpv,gpd},
both quark and gluon contributions are dominated by the imaginary parts
which have the same sign, {\it i.e.} $\varphi_{qg}\simeq 0$.
In the present calculation $\varphi_{qg}\simeq 30^{\circ}$ is obtained.
Considering the wave functions of $\rho^0$
and $\phi$ mesons to be similar (as it is supported by the measured
values of their decay widths), $C_{\rho^0}\simeq C_{\phi}$
follows, and the ratio
of $\phi$ to $\rho^0$ cross sections reads:
\begin{equation}
\frac{\sigma_{\phi}}{\sigma_{\rho^0}}=\frac{2}{9}
\frac{|{\cal T}_g|^2}{|{\cal T}_q|^2+2|{\cal T}_q|
|{\cal T}_g|\cos(\varphi_{qg})+|{\cal T}_g|^2}.
\label{phitorho}
\end{equation}

At {\sc Hermes}, the ratio of $\sigma_{\phi}/\sigma_{\rho^0}$ was measured
\cite{phitorho}. The experimental value was 0.08$\pm$0.01,
slightly increasing with $Q^2$. Inserting
it to the l.h.s of Eq.~(\ref{phitorho})
and taking $\varphi_{qg}$ = 0$^{\circ}$(30$^{\circ}$)
yields $\frac{|{\cal T}_q|}{|{\cal T}_g|}
\Bigr |_{{Hermes}}$ = 0.7 (0.78). This value is in
good agreement with the results of the present calculation, where
$\frac{|{\cal T}_q|}{|{\cal T}_g|}$ ranges between
0.8 and 0.5 (0.5 and 0.3) for the factorized (Regge) ansatz
when $W$ increases
from 4 to 6 GeV. This is in contrast to the above mentioned result
of a dominant quark contribution, $\frac{|{\cal T}_q|}{|{\cal T}_g|}
\simeq 3$ \cite{gpd,gpv}.
Hence, it is concluded that $H_g$ can not be neglected,
{\it i.e.} to arrive at the measured cross section
both quark and gluon amplitudes have to be scaled down in a similar
proportion.

\subsection{Expected Value of TTSA and Projected Statistical Uncertainty}
\label{ttsarhoexp}
The $A_{UT}^{\sin(\phi-\phi_S)}$ amplitude of the TTSA
in Eq.~(\ref{ttrasyform}) is calculated at {\sc Hermes} kinematics.
The statistical error is extrapolated from a preliminary analysis of the {\sc Hermes}
longitudinal target-spin asymmetry measured on the deuteron
\cite{ulrike} that is based on
8~million DIS events. In the latter analysis the data is not split into 
parts corresponding to longitudinal and transverse virtual photons,
while the present calculation is related to longitudinal photons only.
At {\sc Hermes} kinematics ($\langle Q^2 \rangle \simeq 2$~GeV$^2$), 
longitudinal photons constitute about 50\% of 
all virtual photons.
Also, the transverse target polarization is 0.75 while the
longitudinal one is 0.85.
The projected statistical error for 8~million
DIS events taken
on a transversely polarized target is then larger by a
factor $\sqrt{2}\frac{0.85}{0.75}=1.6$ compared to that of Ref.~\cite{ulrike}.
Note that this error estimate may be considered
`optimistic', since it assumes that the contribution to the asymmetry
from longitudinal and transverse photons can be completely
disentangled.

\begin{figure}
\centering
\begin{minipage}[c]{0.49\hsize}
\epsfig{file=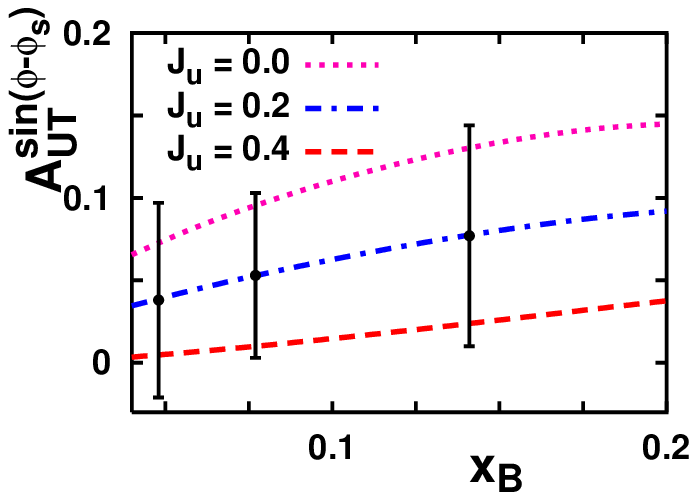,width=\hsize}
\end{minipage}
\begin{minipage}[c]{0.49\hsize}
\epsfig{file=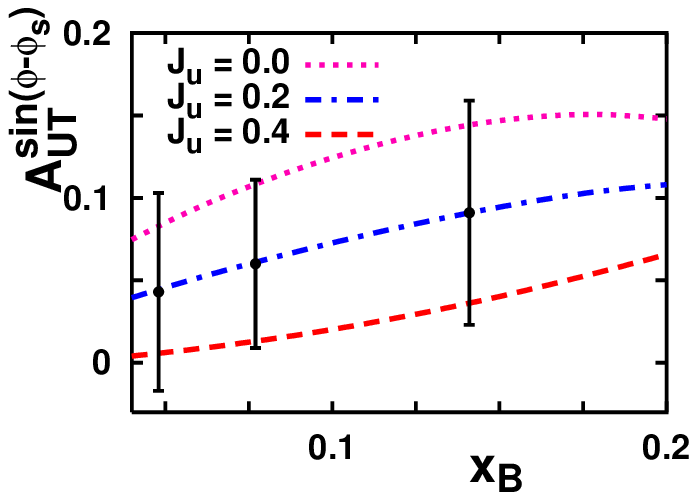,width=\hsize}
\end{minipage}
\\
\begin{minipage}[c]{0.49\hsize}
\epsfig{file=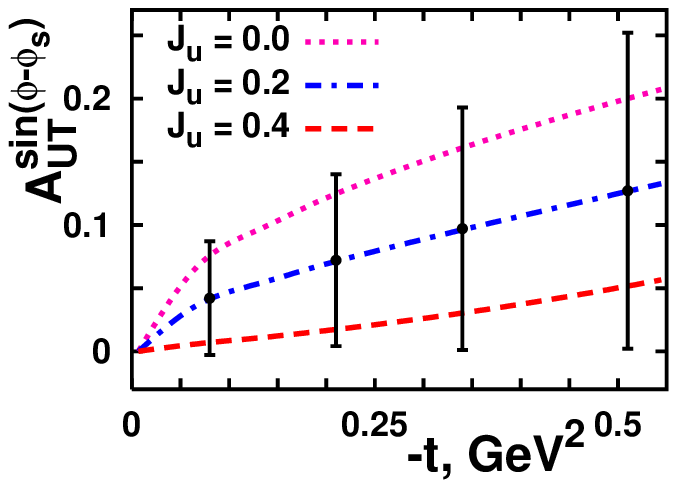,width=\hsize}
\end{minipage}
\begin{minipage}[c]{0.49\hsize}
\epsfig{file=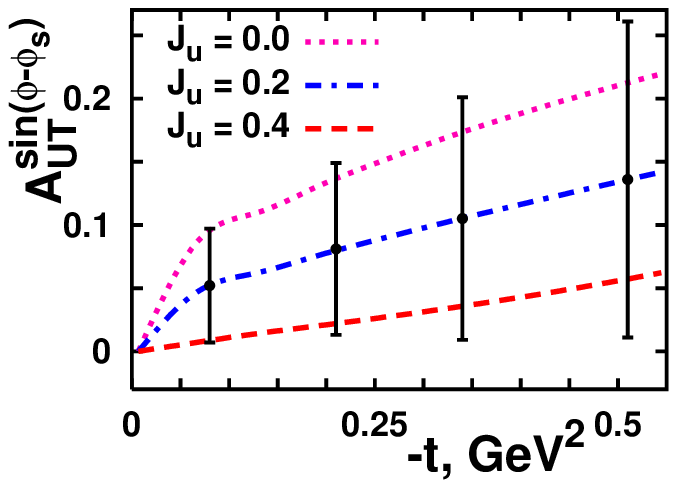,width=\hsize}
\end{minipage}
\caption{Comparison of expected $A_{UT}^{\sin(\phi-\phi_s)}$ amplitudes of the
$\rho^0$ TTSAs
calculated in the Regge ansatz with $b_{val}=1$ and $b_{sea}=1$ 
(left) or $b_{sea}=\infty$ (right). Average kinematic values 
$\langle -t \rangle$ = 0.14 GeV$^2$ and $\langle x_B \rangle $ = 0.085 
for $x_B$ and $t$-dependences, respectively, and $\langle Q^2 \rangle =
2$~GeV$^2$ correspond to a preliminary analysis 
of {\sc Hermes} data on a longitudinally polarized deuterium
target \cite{ulrike}. 
Projected statistical errors are shown.
The systematic uncertainty is expected to be smaller
than the statistical one.}
\label{figtsax}
\end{figure}

\begin{figure}
\centering
\begin{minipage}[c]{0.49\hsize}
\epsfig{file=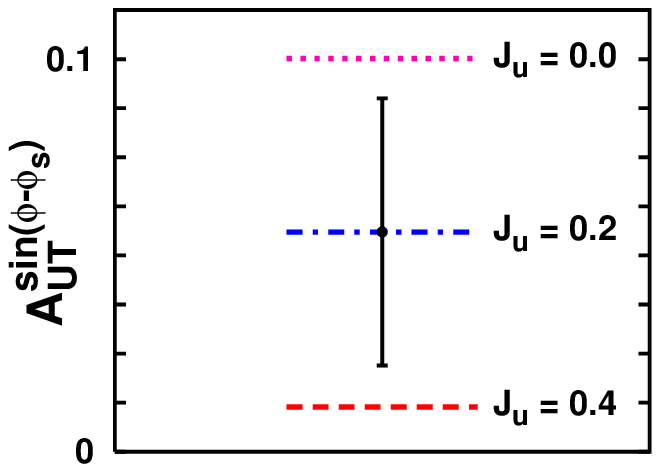,width=\hsize}
\end{minipage}
\begin{minipage}[c]{0.49\hsize}
\epsfig{file=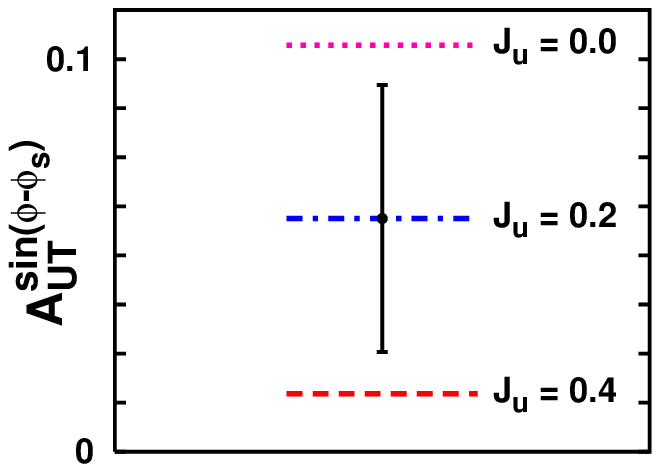,width=\hsize}
\end{minipage}
\caption{Comparison of expected $A_{UT}^{\sin(\phi-\phi_s)}$ amplitudes of the
$\rho^0$ TTSAs calculated 
at average {\sc Hermes} kinematics ($\langle -t \rangle$ = 0.14 GeV$^2$, $\langle x_B \rangle $ = 0.085, 
$\langle Q^2 \rangle = 2$~GeV$^2$)
in the 
Regge ansatz with $b_{val}=1$ and $b_{sea}=1$ (left) or $b_{sea}=\infty$ 
(right).  
Projected statistical errors are shown.
The systematic uncertainty is expected to be smaller
than the statistical one.}
\label{figrhofin}
\end{figure}

The calculated $x_B$ and $t$-dependences of
$A_{UT}^{\sin(\phi-\phi_S)}$ are shown
in Fig.~\ref{figtsax} for different values of $J_u$.
As in the case of DVCS, $J_d$ is fixed inspired by the fact that the
$d$-quark contribution is still suppressed, although the suppression
in $\rho^0$ production is half as strong as in DVCS.
Again, the choice of $J_d=0$ is based on
the results of recent lattice calculations
(see e.g. Ref.~\cite {lattice_prl_qcdsf}).
Note that in contrast to DVCS, $E=0$ results in a vanishing asymmetry.
As it can be seen from comparing Fig.~\ref{figtsax} to Figs.~\ref{fig:tta8}
and \ref{fig:tta9},
the expected magnitude of
$A_{UT}^{\sin(\phi-\phi_S)}$ in $\rho^0$ production
is much smaller than that in DVCS. This is due to
a large gluonic contribution
to the amplitude, which is considered as ``passive'' ($E_g=0, H_g\ne 0$),
{\it i.e.} the gluons dilute the asymmetry in this case.
It was found that the difference in $A_{UT}^{\sin(\phi-\phi_S)}$
between the factorized and Regge ans\"{a}tze is negligible.
Also the variation of $b_{sea}$ only leads to 
a small difference as can be seen
when comparing the left and right panels of
Fig.~\ref{figtsax}, where $x_B$- and $t$-dependences
of the $A_{UT}^{\sin(\phi-\phi_S)}$
amplitude of the asymmetry are shown for $b_{sea}=1$ and $b_{sea}=\infty$,
respectively. The amplitude of the integrated TTSA
is shown in Fig.~\ref{figrhofin}, for the same two cases. It is essentially
independent of $b_{sea}$ and
ranges 
between values of $0.10$ and $0.01$ when $J_u$ ranges 
between zero and 0.4. 

The projected statistical error for 
the integrated TTSA amplitudes is 0.034. Extrapolating the knowledge
on the systematic uncertainty from Ref.~\cite{ulrike},
its size can be expected to be about 0.02,
such that a total experimental uncertainty
below 0.04 appears as a realistic estimate. 
Altogether, the difference in $A_{UT}^{\sin(\phi-\phi_S)}$
due to a change of $J_u$
between zero and 0.4 corresponds to an about 2$\sigma$ effect, where $\sigma$
denotes the {\em total} experimental uncertainty. 
Thus it can be expected that the upcoming $\rho^0$ electroproduction
measurements
performed at {\sc Hermes} will provide an additional
constraint on the size of $J_u$.

A tempting possibility provided by $\rho^0$ production is related
to an estimate of the gluonic content of $E$.
Strongly simplifying, Eq.~(\ref{ttrasyform}) represents the ratio
$E/H \propto (E_q+E_g)/(H_q+H_g)$. Hence, when comparing 
the earlier calculations \cite{gpv} where gluons have been neglected
($E_g=H_g=0$) to the case of ``passive'' gluons presented above
($E_g=0$, $H_g\ne 0$), the asymmetry gets 
smaller (`diluted') by the presence of the term containing $H_g$ in 
the denominator.
On the other hand, if the measured asymmetry would be found large, this could
imply that the gluons are ``active'' ($E_g\ne 0$), so that their
contribution to the spin-flip amplitude can not be neglected.

\section{Summary and Outlook}
\label{summary}
Transverse target-spin asymmetries (TTSAs) in DVCS and $\rho^0$
elastic electroproduction
are the only candidates known by now to access the GPD $E$ on a proton target,
in which $E$ comes as a leading term. A code \cite{vgg} based 
on the model developed in Ref.~\cite{gpd,gpv} is used to calculate 
the expected 
TTSAs to be measured in DVCS on the {\sc Hermes} 
transversely polarized hydrogen 
target. To check the accessibility of $E$ at {\sc Hermes}, different 
parameterization ans\"atze and parameters of $H$ and $E$ are chosen. 
As the model for $E$ depends on the total angular momentum of the 
$u$-quarks in the proton, the possibility arises to check the sensitivity
of the data to different values chosen as $J_u=0.4,0.2,0.0$, while on the
basis of $u$-quark dominance and recent lattice calculations (see
e.g. Ref.~\cite {lattice_prl_qcdsf}) a fixed 
value $J_d=0$ is used.
The calculations are performed at the {\sc Hermes} average kinematic values 
\cite{frank}. The results show that the DVCS TTSA amplitude
$A_{UT}^{\sin{(\phi-\phi_S)}\cos{\phi}}$ is sensitive to 
the GPD $E$ and thus to the total $u$-quark angular momentum $J_u$,
while $A_{UT}^{\cos{(\phi-\phi_S)}\sin{\phi}}$ is not. 
It was found that aside from $J_u$
the amplitude $A_{UT}^{\sin{(\phi-\phi_S)}\cos{\phi}}$
is largely independent on different parameterization
ans\"atze and model parameters.
Projected statistical errors for the asymmetries are 
evaluated by converting the ones from Ref.~\cite{frank} to a data set 
corresponding to 8 million DIS events taken on a transversely polarized 
hydrogen target. 

The same parameterizations are used to calculate the TTSA in
$\rho^0$ electroproduction by longitudinal virtual photons.
The main difference to the DVCS case is a 
large gluonic contribution to the amplitude. At present, only the 
spin-nonflip part of the gluonic amplitude can be reasonably described,
while the spin-flip gluonic GPD $E_g$ is totally unknown. Therefore,
throughout the calculation $E_g$ is set to zero (``passive'' gluons).
Under this assumption, the situation in $\rho^0$
electroproduction appears less favorable concerning the
sensitivity of the expected TTSA amplitude to the
total angular momentum $J_u$.
However, should the value of
the amplitude be measured larger than that predicted by
these calculations, this would imply that $E_g$ can not be neglected,
and thus indicate that
gluons inside the proton carry significant orbital angular momentum.

Altogether, transverse target-spin asymmetries in both DVCS and 
$\rho^0$ electroproduction are studied to evaluate projected uncertainties
for extracting the value of $J_u$ from future data.
Considering all anticipated {\sc Hermes} data to be taken 
for DVCS ($\rho^0$-production), the projected total experimental
$1\sigma$-uncertainty is estimated to correspond to a range of about 
0.1 (0.2) in $J_u$.

\section*{Acknowledgments}
This study would have been impossible without the permanent advice of
M.~Diehl. The support of J.~Volmer is highly appreciated by Z.Y. A.V.
is grateful to A.~Borissov for useful discussions. The advice of
E.~Aschenauer is appreciated. A.V. was supported by the
Alexander von Humboldt foundation, RFBR grants 04-02-16445,
03-02-17291 and Heisenberg-Landau program. This work was supported in part by 
the US Department of Energy.

\section*{Appendix:~TTSA Calculation in DVCS}
\label{ttsadvcscalc}
A code \cite{vgg} is used to estimate the TTSA related to DVCS. 
The coordinate system and angles defined in the code
are the same as depicted in Fig.~\ref{fig:coordinate}. The polarization of the 
target in the code is defined according to
the virtual photon direction. For a transversely polarized target,
the target polarization direction can be chosen either
in the lepton plane ($x$ direction) or perpendicular to
it ($y$ direction). The former corresponds to $\phi_S=0$ or
$\pi$, the latter to $\phi_S=\pi/2$ or
$3\pi/2$. Therefore the following intermediate asymmetries can be calculated:
\begin{eqnarray}
A_x(\phi)&=&\frac{d\sigma_{\phi_S=0}(\phi)-
d\sigma_{\phi_S=\pi}(\phi)}{d\sigma_{\phi_S=0}
(\phi)+d\sigma_{\phi_S=\pi}(\phi)},\nonumber\\
A_y(\phi)&=&\frac{d\sigma_{\phi_S=\frac{\pi}{2}}(\phi)-d\sigma_{\phi_S=\frac{3\pi}{2}}(\phi)}{d\sigma_{\phi_S=\frac{\pi}{2}}(\phi)+d\sigma_{\phi_S=\frac{3\pi}{2}}(\phi)}.
\label{eqn:ref0}
\end{eqnarray}
Defining the following functions
\begin{eqnarray}
A_1(\phi)&=&A_x\cdot\sin{\phi}-A_y\cdot\cos{\phi}, \label{eqn:as12}\\
A_2(\phi)&=&A_x\cdot\cos{\phi}+A_y\cdot\sin{\phi}, \nonumber
\end{eqnarray}
the contribution of the transverse target polarization component of the interference term $\mathcal{I}_{TP}$ to the total cross section in Eq.~(\ref{eqn:cs2}) can be expressed as:
\begin{equation}
d\sigma_{TP}=d\sigma_{unp}\Bigl[ A_1(\phi)\cdot\sin{(\phi-\phi_S)}+A_2(\phi)\cdot\cos{(\phi-\phi_S)}\Bigr].
\end{equation}
Therefore the asymmetries defined in Eq.~(\ref{eqn:tta10}) can be computed as:
\begin{eqnarray}
A^{\sin{(\phi-\phi_S)}\cos{\phi}}_{UT}
&=&A_1^{\cos{\phi}}, \nonumber\\
A^{\cos{(\phi-\phi_S)}\sin{\phi}}_{UT}
&=&A_2^{\sin{\phi}}. \nonumber\\
\nonumber\end{eqnarray}

\end{document}